\documentclass[12pt,a4paper]{article}
\usepackage{color,graphics,epsfig,amsmath,amsfonts,citesort}

\begin{document}


\setlength{\unitlength}{1mm}
\def\noi{\noindent}
\def\lsim{\raise0.3ex\hbox{$<$\kern-0.75em\raise-1.1ex\hbox{$\sim$}}}
\def\gsim{\raise0.3ex\hbox{$>$\kern-0.75em\raise-1.1ex\hbox{$\sim$}}}
\def\bce{\begin{center}}
\def\ece{\end{center}}
\def\bea{\begin{eqnarray}}
\def\eea{\end{eqnarray}}
\def\beq{\begin{equation}}
\def\eeq{\end{equation}}
\def\ba{\begin{array}}
\def\ea{\end{array}}
\def\nn{\nonumber}
\def\micro{{\tt micrOMEGAs}}
\def\calc{{\tt CalcHEP}}
\def\comp{{\tt CompHEP}}
\def\nmh{{\sc nmhdecay}}
\def\susy{{\sc susy}}
\def\ma{m_A}
\def\wh{\widehat}
\def\wt{\widetilde}
\def\a{\alpha}
\def\b{\beta}
\def\d{\delta}
\def\D{\Delta}
\def\e{\epsilon}
\def\g{\gamma}
\def\G{\Gamma}
\def\l{\lambda}
\def\k{\kappa}
\def\t{\theta}
\def\s{\sigma}
\def\S{\Sigma}
\def\x{\chi}
\def\sf{\wt f}
\newcommand{\ch}[1]{\wt \x^\pm_{#1}}
\newcommand{\nt}[1]{\wt \x^0_{#1}}
\newcommand{\mch}[1]{m_{\wt \x^\pm_{#1}}}
\newcommand{\mnt}[1]{m_{\wt \x^0_{#1}}}
\def\bino{\wt B}
\def\wino{\wt W}
\def\higgsino{\wt H}
\def\singlino{\wt S}
\def\gev{\mathrm{GeV}}
\def\tev{\mathrm{TeV}}
\def\tb{\tan\!\b}
\def\ctb{\cot\!\b}
\def\cb{\cos\!\b}
\def\sb{\sin\!\b}
\def\sbb{\sin\!2\b}
\def\tw{\tan\!\t_W}
\def\cw{\cos\!\t_W}
\def\sw{\sin\!\t_W}
\def\Oh2{\Omega h^2}
\renewcommand{\theequation}{\thesection.\arabic{equation}}
\makeatletter
\@addtoreset{equation}{section}
\makeatother


\bce
{\Large\bf Relic density of dark matter in the NMSSM} \\[10mm]
{\large G.~B\'elanger$^1$, F.~Boudjema$^1$, C.~Hugonie$^2$, A.~Pukhov$^3$,
A.~Semenov$^4$} \\[4mm]
{\it
1) Laboratoire de Physique Th\'eorique LAPTH, F-74941 Annecy-le-Vieux, France \\
2) Laboratoire de Physique Theorique, Univ. Paris XI, F-91405 Orsay, France \\
3) Skobeltsyn Inst. of Nuclear Physics, Moscow State Univ., 119992 Moscow,
Russia \\
4) Joint Inst. for Nuclear Research, 141980 Dubna, Russia
}\\[4mm]
\today
\ece

\begin{abstract}
We present a code to compute the relic density of dark matter in the
Next-to-Minimal Supersymmetric Standard Model (NMSSM). Dominant corrections to
the Higgs masses are calculated with \nmh\ as well as theoretical and collider
constraints. All neutralino annihilation and coannihilation processes are then
computed with an extended version of \micro, taking into acount higher order
corrections to Higgs vertices. We explore the parameter space of the NMSSM and
consider in particular the case of a bino LSP, of a mixed bino-higgsino LSP and
of a singlino LSP. As compared to the MSSM, neutralino annihilation is often
more efficient as it can take place via (additional) Higgs resonances as well as
annihilation into light Higgs states. Models with a large singlino component can
be compatible with WMAP constraints.
\end{abstract}

\section{Introduction}

In any supersymmetric (\susy) extension of the Standard Model (SM) with
conserved R-parity, the lightest \susy\ particle (LSP) constitutes a good
candidate for cold dark matter. Recent measurements from WMAP~\cite{WMAP} have
constrained the value for the relic density of dark matter within 10\% ($.0945 <
\Oh2 < .1287$ at $2\s$). The forthcoming PLANCK experiment should reduce this
interval down to 2\%. It is therefore of the utmost importance to calculate the
relic density as accurately as possible in any given \susy\ model, in order to
match this experimental accuracy. This has been done in the context of the
Minimal Supersymmetric Standard Model (MSSM), with the publicly available
program \micro~\cite{micrOMEGAs}. This code computes the relic density of the
lightest neutralino LSP by evaluating the thermally averaged cross section for
its annihilation as well as, when necessary, for its coannihilation with other
\susy\ particles. It then solves the density evolution equation numerically,
without using the freeze-out approximation.

The impact of the WMAP constraints on the parameter space of the MSSM has been
widely studied~\cite{WMAPSUSY}. In the mSUGRA inspired version of the MSSM, one
needs rather special conditions among parameters to have a large enough
annihilation cross-section and meet the constraints of WMAP. This is because the
LSP is usually a bino and the main mechanism for annihilation is into fermion
pairs and requires the presence of a light slepton. Given the direct constraints
on light sleptons and on light neutralinos as well as on the light Higgs mass,
this possibility is valid into a small corner of parameter space. Within mSUGRA
models, to satisfy the WMAP constraint one must then appeal to specific
processes such as coannihilation processes, rapid annihilation through Higgs
exchange or some non negligible higgsino component of the neutralino LSP. The
latter condition is satisfied in the so-called focus point region of mSUGRA
models. On the other hand, in the general MSSM with no assumption on GUT scale
physic and free parameters at the electroweak (EW) scale, it is much easier to
satisfy the upper bound on relic density from WMAP. All models with a mixed
bino/higgsino or mixed bino/wino LSP annihilate much more efficiently and this
even for heavy sfermions. In fact the annihilation is so efficient that the
relic density often lies below the WMAP range, thus implying some additional
dark matter component. Large higgsino component of the neutralino LSP means
enhanced couplings to the $Z$ boson and to the pseudo-scalar Higgs boson $A$,
thus favouring annihilation channels into fermion pairs. Large wino component
means more efficient couplings to gauge bosons favouring annihilation channels
into W pairs. Furthermore coannihilation channels with a heavier neutralino
and/or chargino are also enhanced for either higgsino or wino-like LSP's. In the
limit where all sfermions are heavy the important parameters that enter the
computation are those of the neutralino sector $M_1,M_2,\mu,\tb$ as well as the
pseudo-scalar Higgs mass $\ma$. The trilinear couplings $A_{\sf}$ can play a
role for the relic density as they influence the mass of the lightest Higgs.

It is well known, however, that the MSSM faces a naturalness problem -- the
so-called $\mu$ problem~\cite{mupb} --  arising from the presence of a \susy\
conserving mass term for the Higgs fields in the superpotential, $\mu \wh{H_u}
\wh{H_d}$. The only natural values for the $\mu$ parameter are either zero or
the Planck scale. The first is experimentally excluded while the second
reintroduces the hierarchy problem. The Next-to-Minimal Supersymmetric Standard
Model (NMSSM)~\cite{NMSSM1} provides an elegant solution to the $\mu$ problem
via the introduction of a gauge singlet superfield $\wh{S}$ in the Higgs sector.
Assuming the simplest possible scale invariant form of the superpotential, which
contains the dimensionless $\l\wh{S}\wh{H_u}\wh{H_d}$ coupling, the scalar
component $S$ of the singlet acquires naturally a vacuum expectation value (vev)
of the order of the \susy\ breaking scale, giving rise to an effective
$\mu\equiv\l \langle S \rangle$ of order the EW scale. The NMSSM is the simplest
\susy\ extension of the SM in which the EW scale originates from the \susy\
breaking scale {\em only}. Another nice feature of the NMSSM is that the fine
tuning problem originating from the non-observation of a neutral Higgs boson at
LEP, is less severe than in the MSSM~\cite{FINET}. A possible cosmological
domain wall problem \cite{abel1} can be avoided by introducing suitable
non-renormalizable operators \cite{abel2} that do not generate dangerously large
singlet tadpole diagrams \cite{tadp}. In addition to the MSSM fields, the NMSSM
contains an extra scalar and pseudo-scalar neutral Higgs bosons, as well as an
additional neutralino, the singlino. The phenomenology of the model can be
markedly different from the MSSM~\cite{NMSSM2}. The upper bound on the mass of
the lightest Higgs state is larger than in the MSSM, up to 180
GeV~\cite{NHIGGS}. Moreover, a very light Higgs boson (as light as a few GeV) is
not experimentally excluded~\cite{NLHC}. Similarly, a very light neutralino with
a large singlino component may have escaped LEP searches~\cite{NLEP}. All these
properties could impact significantly the relic density computation in the
NMSSM.

Up to now there has been only a few studies of the relic density of dark matter
in the NMSSM~\cite{NRELDEN}, although a detailed analysis of dark matter direct
detection in this model was recently carried out~\cite{NDIRDET}. In this paper
we present a code that calculates the relic density of dark matter in the NMSSM.
This code provides an interface between \nmh\ and \micro. The FORTRAN code \nmh\
allows a precise calculation of the particle spectrum in the NMSSM, as well as a
complete check of all the available experimental constraints from
LEP~\cite{NMHDECAY}. The parameters are then fed into a new version \micro\
extended to the NMSSM, which calculates {\it all} relevant cross-sections for
neutralino annihilations and coannihilations and computes the relic density.

The paper is organized as follows: we first describe the model, summarize the
main features of \nmh\ and discuss the implementation of the NMSSM into
\comp/\calc\ and \micro. In section 3 we give results for typical case studies.
Conclusions are given in section 4.

\section{Overview of the NMSSM and \nmh}

Extending the calculation of the lightest neutralino relic density from the MSSM
to the NMSSM, requires first to compute the neutralino as well as the Higgs
boson masses and mixings in this model. To achieve this task, we used the
FORTRAN program \nmh~\cite{NMHDECAY}, which conventions we will review now. We
then discuss the implementation of the model into \comp/\calc\ and \micro, as
well as the interface between both codes. We finally give a summary of the
experimental and theoretical constraints taken into account.

\subsection{General Set Up}

In addition to the standard MSSM Yukawa couplings for quarks and leptons, the
NMSSM superpotential contains two additional terms involving the Higgs doublet
superfields, $\wh{H_u}$ and $\wh{H_d}$, and the gauge singlet $\wh{S}$
\beq
W= \l\wh{S}\wh{H_u}\wh{H_d} + \frac{1}{3}\k\wh{S}^3 \,.
\label{WNMSSM}
\eeq
As mentioned in the introduction, the superpotential in eq.~(\ref{WNMSSM}) is
scale invariant, and the EW scale appears only through the soft SUSY breaking
terms in ${\cal L}_\mathrm{soft}$, which in our conventions are given by
\bea
-{\cal L}_\mathrm{soft} & = & m_\mathrm{H_u}^2|H_u|^2 + m_\mathrm{H_d}^2|H_d|^2
+ m_\mathrm{S}^2 | S |^2 \nn\\
& & + (\l A_\l H_uH_dS + \frac{1}{3}\k A_\k S^3 + \mathrm{h.c.}) \nn\\
& & - \frac{1}{2}\left(M_2\l_2\l_2 + M_1\l_1\l_1 + \mathrm{h.c.}\right) \,.
\eea
We denote the Higgs vevs $v_u$, $v_d$ and $s$ such that
\beq
H_u^0 = v_u + \frac{H_{uR} + iH_{uI}}{\sqrt{2}} \,,\quad
H_d^0 = v_d + \frac{H_{dR} + iH_{dI}}{\sqrt{2}} \,,\quad
S = s + \frac{S_R + iS_I}{\sqrt{2}} \,.
\eeq
One can derive three minimization conditions for the Higgs vevs and use them to
re-express the soft breaking Higgs masses in terms of $\l$, $\k$, $A_\l$,
$A_\k$, $v_u$, $v_d$ and $s$. It is also convenient to define the quantities
\beq
\mu = \l s \,, \quad \nu = \k s \,, \quad \tb = \frac{v_u}{v_d} \quad
\mathrm{and} \quad v^2 = v_u^2 + v_d^2 \,.
\eeq

\subsection{Higgs Sector}

In the basis $(H_{uR}, H_{dR}, S_R)$ one obtains the following mass matrix for
the neutral scalar Higgs states:
\bea
{\cal M}_{S,11}^2 & = & M_Z^2\sin^2\!\b + \mu\ctb(A_\l+\nu) \,,\nn\\
{\cal M}_{S,22}^2 & = & M_Z^2\cos^2\!\b + \mu\tb(A_\l+\nu) \,,\nn\\
{\cal M}_{S,12}^2 & = & \left(\l^2v^2-\frac{M_Z^2}{2}\right)\sbb - \mu(A_\l+\nu)
\,,\nn\\
{\cal M}_{S,33}^2 & = & \frac{\l^2v^2A_\l\sbb}{2\mu}\, + \nu(A_\k+4\nu) \,,\nn\\
{\cal M}_{S,13}^2 & = & \l v(2\mu\sb-(A_\l+2\nu)\cb) \,,\nn\\
{\cal M}_{S,23}^2 & = & \l v(2\mu\cb-(A_\l+2\nu)\sb) \,.
\label{scalaire}
\eea
After diagonalization by an orthogonal $3 \times 3$ matrix $S_{ij}$ one obtains
3 neutral scalars $h_i$ (ordered in mass). Similarly, in the basis $(H_{uI},
H_{dI}, S_I)$, the neutral pseudo-scalar mass matrix reads
\bea
{\cal M}_{P,11}^2 & = & \mu\ctb(A_\l+\nu) \,,\nn\\
{\cal M}_{P,22}^2 & = & \mu\tb(A_\l+\nu) \,,\nn\\
{\cal M}_{P,12}^2 & = & \mu(A_\l+\nu) \,, \nn\\
{\cal M}_{P,33}^2 & = & \frac{\l^2v^2\sbb}{2\mu}\, (A_\l+4\nu) - 3A_\k\nu
\,,\nn\\
{\cal M}_{P,13}^2 & = & \l v\cb(A_\l-2\nu) \,,\nn\\
{\cal M}_{P,23}^2 & = & \l v\sb(A_\l-2\nu) \,.
\label{pseudoscalaire}
\eea
Eliminating the Goldstone mode by the rotation:
\beq
\left(\ba{c}H_{uI} \\  H_{dI} \\ S_I \ea\right) = 
\left(\ba{ccc} \cb & -\sb & 0 \\ \sb & \cb & 0 \\ 0 & 0 & 1 \ea\right)
\left(\ba{c} A \\ G \\ S_I \ea\right)
\eeq
the $2 \times 2$ pseudo-scalar mass matrix in the basis ($A, S_I$) has the
following matrix elements
\bea
{\cal M'}_{P,11}^2 & = & \frac{2\mu}{\sbb}\, (A_\l+\nu) \,,\nn\\
{\cal M'}_{P,22}^2 & = & \frac{\l^2v^2\sbb}{2\mu}\, (A_\l+4\nu) - 3A_\k\nu
\,,\nn\\
{\cal M'}_{P,12}^2 & = & \l v (A_\l-2\nu) \,.
\eea
It can be diagonalized by an orthogonal $2 \times 2$ matrix $P'_{ij}$, yielding
2 neutral pseudo-scalars $a_i$ (ordered in mass). Finally, the charged Higgs
mass is given by
\beq
m_{h^\pm}^2 = \frac{2\mu}{\sbb}\, (A_\l+\nu) + M_W^2 - \l^2v^2 \,.
\label{charge}
\eeq

In the MSSM limit ($\l \to 0$, $\mu$ fixed) one obtains 2 quasi pure singlet
states with masses
\beq
m_S^2 = \nu(A_\k+4\nu) \,, \quad m_P^2 = -3A_\k\nu \,.
\label{mSP}
\eeq
In the doublet sector, one can define
\beq
m_A^2 = {\cal M'}_{P,11}^2 = \frac{2\mu}{\sbb}\, (A_\l+\nu) \,.
\label{mA}
\eeq
In the limit
$m_A \gg M_Z$, the masses of one scalar, one pseudo-scalar and the charged Higgs
states are $\approx m_A$. The mass of the lightest scalar Higgs is bounded by
\beq
m_{h_1}^2 \leq M_Z^2\cos^2\!2\b + \l^2v^2\sin^2\!2\b \,.
\label{mhtree}
\eeq

In order to calculate the relic density, one needs the Feynman rules for triple
Higgs interactions, in case the LSP annihilates into a Higgs pair through
s-channel Higgs boson exchange. We give here the coupling of a non-singlet
scalar $h_i$ to 2 quasi pure singlet pseudo-scalars $P$:
\beq
g_{h_iPP} = \sqrt{2}\l^2 v\, (S_{i1}\sb + S_{i2}\cb) + \sqrt{2}\l\k v\,
(S_{i2}\sb + S_{i1}\cb) \,.
\label{haacoup}
\eeq
The complete formulae for the triple Higgs interactions can be found in
ref.~\cite{NMHDECAY}.

\subsection{Neutralino Sector}

In the basis $(\bino,\wino,\higgsino_u,\higgsino_d,\singlino)$ the $5 \times 5$
neutralino mass matrix reads
{\footnotesize \beq
{\cal M}_{\wt \x^0} = \left( \ba{ccccc}
M_1 & 0 & M_Z\sw\sb & -M_Z\sw\cb & 0 \\
0 & M_2 & -M_Z\cw\sb & M_Z\cw\cb & 0 \\
M_Z\sw\sb & -M_Z\cw\sb & 0 & -\mu & -\l v\cb \\
-M_Z\sw\cb & M_Z\cw\cb & -\mu & 0 & -\l v\sb \\
0 & 0 & -\l v\cb & -\l v\sb & 2\nu
\ea \right) \,.
\label{neumatrix}
\eeq}
This matrix is diagonalized by a unitary matrix, $N$. The lightest neutralino
LSP can then be decomposed as
\beq
\nt 1 = N_{11}\wt B + N_{12}\wt W + N_{13}\wt H_u + N_{14}\wt H_d + N_{15}\wt S
\,.
\eeq
Further, we shall talk about the bino, higgsino and singlino fractions of the
LSP, which we define as $N_{11}^2$, $N_{13}^2+N_{14}^2$ and $N_{15}^2$
respectively.

In the limit $\l \to 0$, the singlino is a quasi pure state with mass
\beq
m_{\singlino} = 2\nu \,.
\label{singmass}
\eeq

For the relic density calculation, the coupling of the LSP to the $Z$ boson is
relevant. This coupling depends only on the higgsino components of the LSP and
is proportional to $N_{13}^2 - N_{14}^2$. The couplings of the LSP to the scalar
and pseudo-scalar Higgs states will enter the computation of LSP annihilation
through a Higgs resonance or t-channel annihilation into Higgs pairs. The
Feynman rule for the LSP-scalar-scalar vertex reads
\bea
g_{\nt 1 \nt 1 h_i} & = & g (N_{12}-N_{11}\tw) (S_{i1}N_{13}-S_{i2}N_{14}) \nn\\
& & + \sqrt{2}\l N_{15} (S_{i1} N_{14} + S_{i2} N_{13}) + \sqrt{2} S_{i3} (\l
N_{13}N_{14} - \k N_{15}^2) \,.
\label{neuneuh}
\eea
The first term is equivalent to the $\nt 1 \nt 1 h$ coupling in the MSSM by
replacing $S_{11}=S_{22}=\cos\!\a$ and $S_{12}=-S_{21}=\sin\!\a$ while the last
two terms are specific of the NMSSM. The second term is proportionnal to the the
singlino component of the LSP while the last one is proportionnal to the singlet
component of the scalar Higgs. The latter involves either the higgsino component
of the LSP or its singlino component. When the LSP has a very small singlino
component, then as in the MSSM, coupling to the scalars requires a mixed
bino/higgsino LSP. Typically in models with a very light scalar Higgs, this one
has an important singlet component and the coupling to the LSP is given by the
last term in eq.~(\ref{neuneuh}). Similarly, the LSP coupling to a
pseudo-scalar reads
\bea
ig_{\nt 1 \nt 1 a_i} & = & g (N_{12}-N_{11}\tw) (P_{i1}N_{13}-P_{i2}N_{14})
\nn\\
& & - \sqrt{2}\l N_{15} (P_{i1} N_{14} + P_{i2} N_{13}) - \sqrt{2} P_{i3} (\l
N_{13}N_{14} - \k N_{15}^2) \,.
\label{neuneua}
\eea

\subsection{Radiative Corrections} \label{sec:rad}

Similarly to the MSSM, radiative corrections to the Higgs masses in the NMSSM
can be relatively large~\cite{NHIGGS}. As said earlier, in order to calculate
the Higgs spectrum, we used the program \nmh~\cite{NMHDECAY}. Let us review now
the accuracy with which radiative corrections are computed for the Higgs sector
in this program. 

First, we assume that the Yukawa couplings and the soft terms are defined at the
\susy\ breaking scale $Q = M_\mathrm{SUSY}$, corresponding to an average of the
squark masses. Quantum fluctuations at higher scales are assumed to be
integrated out through the standard renormalization group evolution of the
parameters from a fundamental scale like $M_\mathrm{GUT}$ down to the scale $Q$.
The effective Lagrangian at the scale $Q$ can be assumed to be of the standard
\susy\ form plus soft terms. The full effective action then reads
\beq
\G_\mathrm{eff} = \sum_i Z_i \ D_{\mu} H_i D^{\mu} H_i - V_\mathrm{eff}(H_i)
\,.
\label{effac}
\eeq
It is obtained from the effective Lagrangian at the scale $Q$ by adding all
quantum effects arising from fluctuations at scales $\lsim Q$. (Here, $H_i$
denotes all the Higgs fields, $H_u$, $H_d$ and $S$.) These quantum effects can
be classified according to powers of the various couplings, and powers of
potentially large logarithms.

Let us start with the corrections to the effective potential. It is somehow more
convenient to classify the corrections to the (lightest) scalar Higgs mass,
which is essentially the second derivative of the effective potential. At tree
level, one can rewrite eq.~(\ref{mhtree}) as
\beq
m_{h_1}^2 \sim (g^2+\l^2) v^2
\eeq
where $g$ denotes the weak gauge couplings (we do not distinguish between large
and small $\tb$ here).

The dominant one loop corrections to $m_{h_1}^2$ originate from top, stop,
bottom and sbottom loops. The corresponding corrections $\delta m_{h_1}^2$ to
$m_{h_1}^2$ are of order
\bea
\delta_t m_{h_1}^2 \sim h_t^4 v^2 \ln\left(Q^2/m_t^2\right)
& , & \delta_{\wt t} m_{h_1}^2 \sim h_t^4 v^2 \,,\nn\\
\delta_b m_{h_1}^2 \sim h_b^4 v^2 \ln\left(Q^2/m_b^2\right)
& , & \delta_{\wt b} m_{h_1}^2 \sim h_b^4 v^2 \,.
\eea
These contributions to the effective potential are computed exactly, without
expanding in large logarithms or squark mass splittings. We also take into
account the one loop pure weak (leading log) contributions of the order $g^4
\ln(Q^2/M_Z^2)$.

The dominant two loop corrections to the effective potential are of the form
\beq
\delta_2 m_{h_1}^2 \sim (h_t^6+h_t^4\alpha_s)v^2\ln^2\left(Q^2/m_t^2\right) \,.
\eeq
Here, only the leading (double) logarithms are included, i.e. we neglect single
logs as well as terms involving bottom and sbottom loops, proportional to powers
of $h_b$.

Next, we review the contributions to the wave function normalization constants
$Z_i$ in eq.~(\ref{effac}). If evaluated for external momenta of ${\cal
O}(m_t)$ (the order of the Higgs masses), top and bottom quark loops
yield contributions to $Z_i$ of the form
\beq
Z_u \sim 1 + h_t^2 \ln\left(Q^2/m_t^2\right) \,, \quad
Z_d \sim 1 + h_b^2 \ln\left(Q^2/m_t^2\right) \,.
\label{Zh}
\eeq
After rescaling the Higgs fields so that their kinetic energies are canonically
normalized, these effects generate contributions $\delta_Z m_{h_1}^2$ which
take the form
\beq
\delta_Z m_{h_1}^2 \sim (g^2+\l^2) (h_t^2+h_b^2) \ln\left(Q^2/m_t^2\right) \,.
\eeq

Once the Higgs mass matrices are diagonalized, one may find eigenstates with
masses $m_H$ substantially larger than $m_t$. The $Z$ factors are then evaluated
at external momenta of ${\cal O}(m_H)$, i.e. at the pole of the propagators.
Hence for $m_H \gg m_t$ the logarithms $\ln(Q^2/m_t^2)$ in eq.~(\ref{Zh}) is
replaced by $\ln(Q^2/m_H^2)$, with coefficients depending on Higgs mixing
matrices.

\subsection{Implementation into \comp/\calc\ and \micro} \label{sec:micro}

The power of \micro~\cite{micrOMEGAs} is that given a set of parameters in the
MSSM or in the NMSSM it first isolates the LSP before generating, for any given
situation including those where coannihilations occur, all the necessary matrix
elements of all relevant processes. Moreover an automatic procedure for looking
for s-channel poles is incorporated into the program such that a more precise
integration routine can be used in the event one is close to a pole. This way,
any kind of annihilation or coannihilation that may be imagined is readily dealt
with. This is possible thanks to the high level of automation based on the
computation of cross sections through \comp/\calc~\cite{COMPHEP}. Moreover an
important advantage is the ease with which a complicated model such as the NMSSM
with its many fields and parameters, and therefore an extremely long list of
defining Feynman rules, can be implemented in this package~\cite{Semenov:prepa}. The gruesome task of
having to code all possible vertices that appear in the NMSSM in order to
generate the matrix elements is taken care of by yet another automatic procedure
that only requires to define the Lagrangian in a very compact form, through
multiplets and superfields. This step is performed by {\tt
LANHEP}~\cite{LANHEP}, a program that generates the complete set of particles
and vertices once given a Lagrangian~\cite{Semenov:2002mssm}. The modification
of \micro\ to go from the MSSM to the NMSSM is done essentially through a
modification of the model file through {\tt LANHEP}.

One drawback, though, is that \comp/\calc\ only deals with tree-level matrix
elements while for the NMSSM, some parameters, especially in the Higgs sector
(notably the lightest Higgs mass and coupling to $b\bar b$) receive important
radiative corrections. For the code to be of any use, these important radiative
corrections need to be taken into account. They are introduced in a gauge
invariant and consistent way through an effective Lagrangian approach. For the
case at hand, to parametrize the radiative corrections to the Higgs masses and
couplings, we write a general dimension four CP conserving effective scalar
potential that involves the two Higgs doublets $H_u$, $H_d$ and the singlet $S$
as
\bea
V_\mathrm{rad} & = & \l_1 (H_u H_u^*)^2/2 + \l_2 (H_d H_d^*)^2/2
+ \l_3 (H_u H_u^*) (H_d H_d^*) \nn\\
& + & \l_4 (\e H_u H_d) (\e H_u^* H_d^*)
+ \l_5 ((\e H_u H_d)^2+(\e H_u^* H_d^*)^2)/2 \nn\\
& + & \l^s_1 (H_u H_u^*) S S^* + \l^s_2 (H_d H_d^*) S S^*
+ \l^s_s (S S^*)^2/2 \nn\\
& + & \l^s_5 ((\e H_u H_d) S^2+(\e H_u^* H_d^*) S^{*2})/2
+ \l^s_p (S^4+S^{*4}) \,.
\eea
With this effective potential, corrections to both the scalar/pseudo-scalar
masses and the trilinear and the quartic Higgs vertices can be expressed in
terms of the ten $\l$'s. The Higgs mass matrices
eqs.~(\ref{scalaire},\ref{pseudoscalaire}) can be rewritten as
\bea
{\cal M}^2_{S,11} & \to & {\cal M}^2_{S,11} + 2 \l_1\ v^2 \sin^2\!\b
- \frac{\l^s_5}{2}\ s^2 \ctb \,,\nn\\
{\cal M}^2_{S,22} & \to & {\cal M}^2_{S,22} + 2 \l_2\ v^2 \cos^2\!\b
- \frac{\l^s_5}{2}\ s^2 \tb \,,\nn\\
{\cal M}^2_{S,12} & \to & {\cal M}^2_{S,12} + (\l_3+\l_4+\l_5)\ v^2 \sbb
+ \frac{\l^s_5}{2}\ s^2 \,,\nn\\
{\cal M}^2_{S,33} & \to & {\cal M}^2_{S,33} + 2 (\l^s_s+4\l^s_p)\ s^2 \,,\nn\\
{\cal M}^2_{S,13} & \to & {\cal M}^2_{S,13} + 2 \l_1^s\ v s \sb + \l_5^s\ v s
\cb \,,\nn\\
{\cal M}^2_{S,23} & \to & {\cal M}^2_{S,23} + 2 \l_2^s\ v s \cb + \l_5^s\ v s
\sb
\eea
and
\bea
{\cal M}^2_{P,11} & \to & {\cal M}^2_{P,11} - 2 \l_5\ v^2 \cos^2\!\b
- \frac{\l_5^s}{2}\ s^2 \ctb \,,\nn\\
{\cal M}^2_{P,22} & \to & {\cal M}^2_{P,22} - 2 \l_5\ v^2 \sin^2\!\b
- \frac{\l_5^s}{2}\ s^2 \tb \,,\nn\\
{\cal M}^2_{P,12} & \to & {\cal M}^2_{P,12} - \l_5\ v^2 \sbb \,,\\
{\cal M}^2_{P,33} & \to & {\cal M}^2_{P,33} - \l^s_5\ v^2 \sbb - 16 \l^s_p\ s^2
\,,\nn\\
{\cal M}^2_{P,13} & \to & {\cal M}^2_{P,13} - \l^s_5\ v s \cb \,,\nn\\
{\cal M}^2_{P,23} & \to & {\cal M}^2_{P,23} - \l^s_5\ v s \sb \,.
\eea
Finally, the charged Higgs mass eq.~(\ref{charge}) can be rewritten as 
\beq
m_{h^\pm}^2 \to m_{h^+}^2 - (\l_4+\l_5)\ v^2 + \frac{\l_5^s\ s^2}{\sbb} \,.
\eeq

Starting from the results for the corrected Higgs masses and mixing angles
provided by \nmh, we then solve for the $\l$'s. Having extracted these
parameters {\tt LANHEP} readily derives the corresponding scalar trilinear and
quartic Higgs vertices. This is an extension of the procedure that was shown in
detail in~\cite{hhh-fb-as}. In the present implementation the full corrections
from \nmh\ to the Higgs masses are included. Consequently the Higgs
self-couplings are effectively corrected at the same leading order. As concerns
these self-couplings, the results differ somehow from those one would obtain
from \nmh. This is due to the fact that, in \nmh, only the one loop leading
logarithms coming from third generation quarks/squarks loops are taken into
account in the Higgs self couplings. This said the vertex which receives the
largest correction, that is the one involving the lightest scalar $h_1h_1h_1$,
does not contribute significantly to neutralino pair annihilation. On the other
hand, as we will see next, the $h_2a_1a_1$ vertex often enters processes of
neutralino annhilation. For these vertices higher order corrections typically do
not exceed a few percent when expressed in terms of corrected masses.

\subsection{Parameter Space Handling}

At the EW scale, the free parameters in the Higgs sector are (at tree level):
$\l$, $\k$, $m_{H_u}^2$, $m_{H_d}^2$,  $m_{S}^2$, $A_\l$ and $A_\k$. Using the
three minimization conditions of the Higgs potential, one can eliminate the soft
Higgs masses in favour of $M_Z, \tb$ and $\mu=\l s$. We thus consider as
independent parameters the following set of variables
\beq
\l , \, \k ,\, \tb ,\, \mu ,\, A_\l , \, A_\k  \,.
\eeq
The soft terms in the squark and slepton sector (which enter the radiative
corrections in the Higgs sector) are also fixed at the EW scale. Since the
gaugino soft masses $M_1$ and $M_2$ enter the neutralino mass matrix, we will
keep them as free parameters in our analysis.

We set all these parameters in the program \nmh~\cite{NMHDECAY}. For each point
in the parameter space, the program \nmh\ first checks the absence of Landau
singularities for $\l$, $\k$, $h_t$ and $h_b$ below the GUT scale. For
$m_t^\mathrm{pole}= 175 \gev$, this translates into  $\l < .75$, $\k < .65$, and
$1.7 < \tb < 54$. \nmh\ also checks the absence of an unphysical global minimum
of the scalar potential with vanishing Higgs vevs.

\nmh\ then computes scalar, pseudo-scalar and charged Higgs masses and mixings,
taking into account one and two loop radiative corrections as mentioned in
section~\ref{sec:rad}, as well as chargino and neutralino masses and mixings.
Finally, all available experimental constraints from LEP are checked:

\noi 1) In the neutralino sector, we check that the lightest neutralino does not
contribute too much to the invisible $Z$ width ($\G(Z \to \nt 1 \nt 1 ) < 1.76$
MeV if $\mnt 1 < M_Z/2$, and that $\s(e^+e^- \to \nt 1 \nt i ) <
10^{-2}~\mathrm{pb}$ if $\mnt 1 + \mnt i < 209$ GeV ($i>1$) and $\s(e^+e^- \to
\nt i \nt j ) < 10^{-1}~\mathrm{pb}$ if $\mnt i + \mnt j < 209$ GeV ($i,j>1$).
In the chargino sector, we check that the lightest chargino is not too light
($\mch 1 < 103.5$ GeV).

\noi 2) In the charged Higgs sector, we impose $m_{h^\pm} > 78.6$ GeV.

\noi 3) In the neutral Higgs sector, we check the constraints on the production
rates (reduced couplings) $\times$ branching ratios versus the masses, for all
of the CP-even states $h$ and CP-odd states $a$, in all the channels studied at
LEP (cf ref.~\cite{NMHDECAY} for details).

For points which violate either a theoretical constraint (Landau Pole or
unphysical global minimum) or an experimental constraint, a warning is issued by
\nmh. The Higgs masses and mixings calculated in \nmh\ are then fed into \micro.
The input parameters of the NMSSM needed by \micro\ are listed in 
table~\ref{micropar}, while the input parameters of the standard model are specified in 
ref.~\cite{micrOMEGAs}. We assume that the masses of the first two generations
of sfermions are equal.
\micro\ recomputes the masses of charginos, neutralinos and sfermions at
tree-level. In the present version the soft terms for sfermions are used and
masses and mixings are computed at tree-level. It is straightforward to replace
this with some external program (such as \nmh) that calculates all masses for
\susy\ particles, as was done in the case of the MSSM. \micro\ then
calculates the LSP relic density, taking into account all possible annihilation
and coannihilation diagrams. For triple Higgs vertices, the main corrections are
included as outlined in section~\ref{sec:micro}. For Higgs couplings to quarks,
leading QCD corrections are taken into account using the running quark masses.

\begin{table}[htbp]
\bce
\begin{tabular}{|l|l||l|l|}
\hline
name & comment & name & comment \\
\hline
hL    & $\l$    & Mli, i=2,3 & masses of left sleptons \\
hK    & $\k$    & Mri, i=2,3 & masses of right sleptons \\
tb    & $\tb$   & Mqi, i=2,3 & masses of left squarks \\
mu    & $\mu$   & Mui, i=2,3 & masses of right u-squarks \\
hLs   & $A_\l$  & Mdi, i=2,3 & masses of right d-squarks \\
hKs   & $A_\k$  & & \\
MG1    & $M_1$   & Al & $\wt \tau$ trilinear coupling \\
MG2    & $M_2$   & Ab & $\wt b$ trilinear coupling \\
MG3    & $M_3$   & At & $\wt t$ trilinear coupling \\
\hline
\end{tabular}
\ece
\caption{Input parameters for \micro\_nmssm.}
\label{micropar}
\end{table}

\section{Results}

In this section, we discuss constraints on the parameter space of the NMSSM
originating from the WMAP results on dark matter relic density. To avoid having
to deal with a large number of parameters, we assume very heavy sfermions
($m_{\sf}=1$~TeV) and fix the trilinear sfermion mixing $A_{\sf}=1.5$~TeV.
Thus annihilation into fermion pairs through t-channel sfermion exchange and
coannihilation with sfermions are suppressed. In
the gaugino sector, we assume universality at the GUT scale, which at the EW
scale corresponds to $M_2 = 2M_1$ and $M_3 = 3.3M_2$. The parameters of our
model are thus $\l, \k, \mu, \tb, A_\l, A_\k$ and $M_2$. These parameters are
free parameters at the EW scale. For the SM parameters, we assume $\a_s=0.118$,
$m_t^\mathrm{pole}=175$~GeV and $m_b(m_b)=4.24$~GeV.

We concentrate on models which can differ markedly from the MSSM predictions, in
particular models with $\tb \leq 5$ for which annihilation through a Higgs
resonance is marginal in the MSSM. First we study the behaviour of $\Oh2$ as a
function of $\mu$ and $M_2$. Then, picking values for $\mu$, $M_2$ corresponding
to typical cases, we present contour plots in the $\l$, $\k$ plane, where the
difference between the MSSM (which we recover at small $\l$) and the NMSSM
appears explicitly. The parameters $A_\l,$ $A_\k$ are also critical as they
affect the masses of the Higgs states and therefore the regions of parameter
space where rapid annihilation through a Higgs resonance can take place.
Finally, we present some choices of parameters for which the LSP is mainly
singlino and the relic density still agrees with the WMAP constraints.

\subsection{MSSM-like case: dependence in $\mu-M_2$} \label{sec:mum2}

\begin{figure}[htbp]
\begin{center}
\begin{picture}(90,95)
\put(0,-7){\mbox{\epsfig{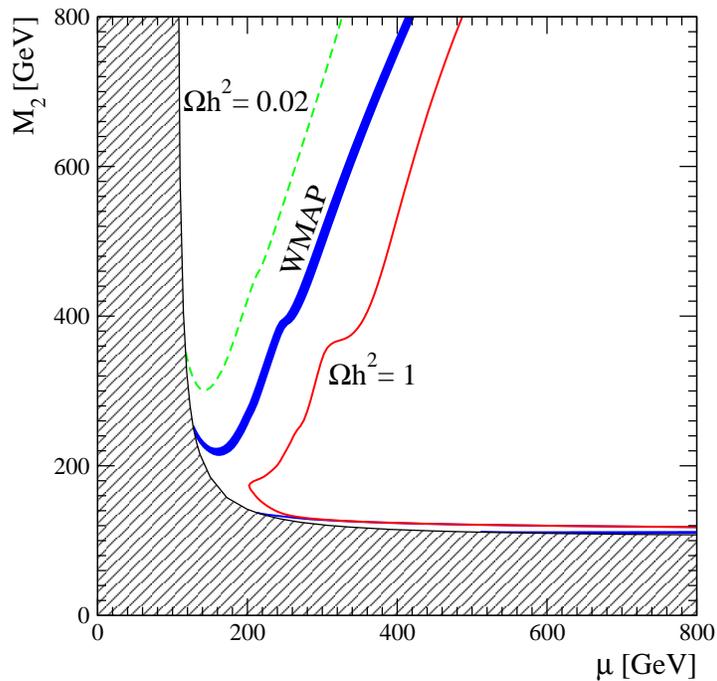}}}
\end{picture}
\end{center}
\caption{Contour plots for $\Oh2 = 0.02$, $0.0945 < \Oh2 < 0.1287$ (WMAP
constraint) and $\Oh2 = 1$ in the $\mu$, $M_2$ plane for $\l=0.1$, $\k=0.1$,
$\tb=5$, $A_\l=500$~GeV and $A_\k=0$. The hatched region is excluded by LEP
constraints on charginos.}
\label{fig:mum2}
\end{figure}

To give an overview of the behaviour of the relic density in the $\mu$, $M_2$
plane, in fig.~\ref{fig:mum2} we consider a model with $\l=0.1$, $\k=0.1$,
$\tb=5$, $A_\l=500$~GeV, $A_\k=0$. For this choice of parameters the singlino
component of the LSP is small so that apart from the Higgs sector and the heavy
neutralinos, the model is MSSM-like. Since the parameters are not specially
tuned to encounter Higgs resonances one expects the predictions for the relic
density to be rather similar to the MSSM. The LEP exclusion region arises from
the limit on chargino pair production. For this choice of parameters the LEP
limits on the Higgs sector does not play a role. Compatibility with WMAP is
found in two different regions of the $\mu$, $M_2$ plane, similar to the ones
obtained in the MSSM at small to intermediate values of $\tb$.

The first region corresponds to a very thin band whith $\mu \gg M_2$, just at
the boundary of the region ruled out by the LEP constraint on charginos.
Although in this case the LSP is mostly a bino, efficient annihilation is
possible via s-channel scalar Higgs ($h_1$) exchange. However some fine-tuning
is required to adjust the mass of the LSP to half the mass of $h_1$. We will
come back to this bino LSP scenario in section~\ref{sec:bino}

The second region is a band where $\mu\ \gsim\ M_1 = M_2/2$. There, the LSP is
mostly bino with just enough higgsino component to annihilate efficiently into
gauge boson pairs ($WW, ZZ$). This is essentially s-channel annihilation via
scalar Higgs or $Z$ exchange as well as t-channel chargino/neutralino exchange.
The higgsino fraction necessary to obtain $\Oh2 \approx 0.1$ increases with the
LSP mass, from 25\% when $\mnt 1 = 140$~GeV to 50\% when $\mnt 1 = 400$~GeV.
When $\mu$ and $M_2$ are large enough so that $\mnt 1 > m_t$, annihilation into
top quark pairs also contributes significantly. The onset of the top pair
annihilation shows up as a kink on the WMAP allowed band. Finally, note that for
$M_2 < 200$~GeV and $\mu \approx M_2$ the relic density is above WMAP. This is
because on the one hand the LSP bino component is large and on the other hand
the gauge boson pair channel is not kinematically available. In
section~\ref{sec:mixed} we will study this mixed bino/higgsino case in more
details.

To the left of the last region, the LSP is mainly higgsino and annihilation into
gauge boson pairs is efficient. For a higgsino LSP, coannihilation processes
with charginos and neutralinos also contribute significantly. Therefore, the
relic density is very small in this region. However, increasing the singlino
content of the LSP may rise the relic density inside the WMAP allowed range. We
will come back to this possibility in section~\ref{sec:sing}.

To the right of the WMAP allowed band, the LSP bino component is large while no
Higgs state has the appropriate mass for a s-channel resonance (recall that we assume
heavy sfermions, so annihilation to fermions through t-channel sfermion exchange
is suppressed). Hence, for our choice of parameters, the relic density is large
in this region of the $\mu, M_2$ plane. However, for different choices of
parameters, one may find areas where s-channel Higgs resonances bring the relic
density down in this region. We will present such scenarios in
section~\ref{sec:res}.

\begin{figure}[htbp]
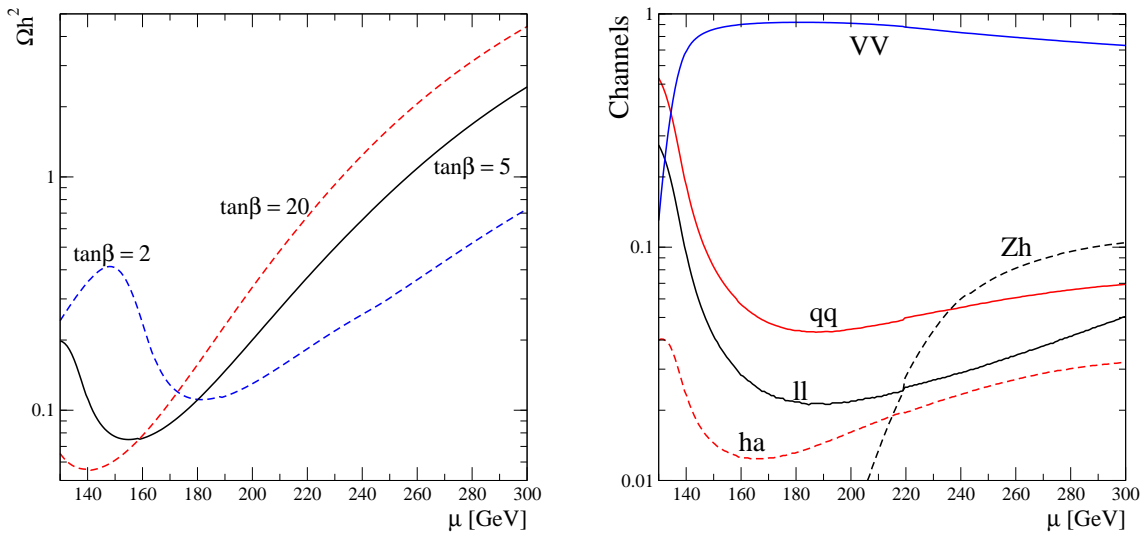

\begin{center}
\begin{picture}(160,75)
\put(0,-7){\mbox{\epsfig{file=fig2a.eps,height=7cm}}}
\put(80,-7){\mbox{\epsfig{file=fig2b.eps,height=7cm}}}
\end{picture}
\end{center}
\caption{a) $\Oh2$ as a function of $\mu$ for $\l=0.1, \k=0.1, A_\l=500$~GeV,
$A_\k=0, M_2=230$~GeV, $\tb=5$ (full) and $\tb=2,20$ (dash). b) Relative
contribution of the main annihilation channels for the case $\tb=5$. Here $VV$
includes both $WW$ and $ZZ$ channels and $qq$ is the sum over all the quarks.}
\label{fig:mu}
\end{figure}

To illustrate more precisely the main mechanisms at work for the LSP
annihilation, we fix $M_2=230$~GeV and plot the relic density as a function of
$\mu$ in fig.~\ref{fig:mu}. The relative contributions of the most important
channels are also displayed. At small values of $\mu$, the LSP mass is below
$M_W$ and the main annihilation channel is into $q\bar{q}$ pairs through $Z$
exchange. The $b\bar{b}$ channel is enhanced as it receives an additional
contribution from $h_1$ exchange. As $\mu$ increases, so does the LSP mass.
Annihilation into gauge boson pairs rises sharply and the relic density drops.
The $WW$ mode is typically 4 times larger than the $ZZ$ mode. For larger $\mu$,
the LSP becomes less higgsino and the relic density increases (recall that the
LSP coupling to gauge bosons depends only on its higgsino components). A
subdominant contribution arises from the $h_1a_1$ channel. This mode is
kinematically accesssible over the whole region probed since the scalar mass is
$m_{h_1} \approx 118$~GeV, the pseudo-scalar mass is $m_{a_1} \approx 20$~GeV and
the LSP mass range is $75 < \mnt 1 < 109$~GeV. For $\mu > 200$~GeV, the $Zh_1$
annihilation channel becomes kinematically accessible. While this channel
accounts for up to $10\%$ of all annihilations, this contribution is not
sufficient to bring the relic density within the WMAP range.

In fig.~\ref{fig:mu}a we also show the relic density for $\tb = 2, 20$. Changing
$\tb$ affects the LSP mass as well as its bino/higgsino fraction, thus having a
large impact on the relic density. First, the LSP mass increases with $\tb$.
This explains why at small $\mu$ (where one can be below the WW threshold) the
relic density is smaller at large $\tb$. On the other hand, the bino component
of the LSP increases with $\tb$. Hence, for $\mu \gsim 180$~GeV (where the WW
channel is kinematically open for all values of $\tb$) annihilation is more
efficient at low values of $\tb$.

\subsection{Bino LSP, annihilation through $h_1$} \label{sec:bino}

\begin{figure}[htbp]
\bce
\begin{picture}(90,95)
\put(0,-7){\mbox{\epsfig{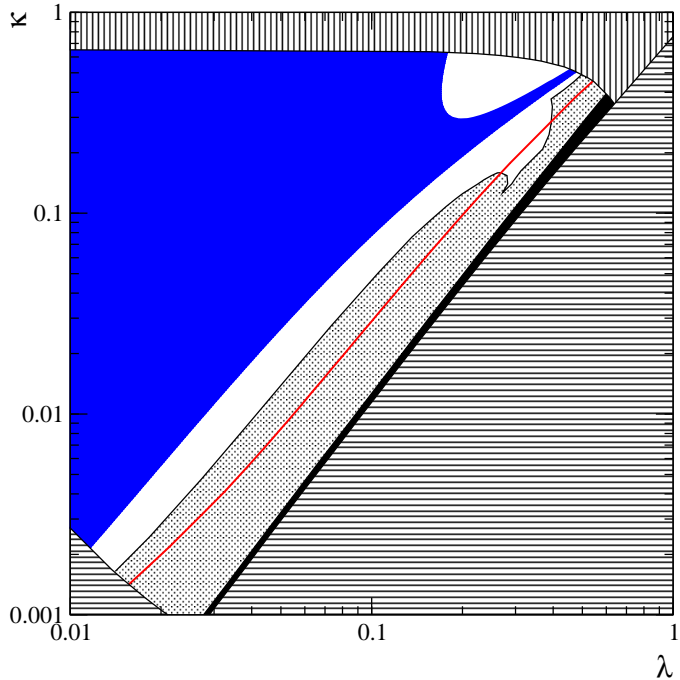}}}
\end{picture}
\ece
\caption{Region in the $\l, \k$ plane for which $0.0945 < \Oh2 < 0.1287$  (blue)
for $\mu=700$~GeV, $M_2=111$~GeV, $A_\l=500$~GeV, $A_\k=0$, $\tb=5$ and  contour
curve for $\Oh2=1$ (red). The theoretically/experimentally excluded regions are
also displayed: LEP Higgs exclusion (grey), Landau pole (vertical lines),
negative Higgs mass squared (horizontal lines) and non-physical global minima of
the scalar potential (black).}
\label{fig:mu700m110_lk}
\end{figure}

If $\mu \gg M_2$, the LSP is almost a pure bino (more than 99\%), and the relic
density is large (t-channel sfermion exchange is suppressed for heavy sfermions
and annihilation through s-channel Z requires some higgsino component)
unless annihilation proceeds through a s-channel Higgs resonance. In
this case, the relic density is very sensitive to the mass
difference, $m_{h_1}-2 m_{\chi_1}$~\cite{Allanach,Kraml}. For $\l=\k=0.1$, $\tb=5$, $A_\l=500$~GeV,
$A_\k=0$, $\mu=700$~GeV, and $M_2=111$~GeV (corresponding to the first WMAP
allowed region in fig.~\ref{fig:mum2}) we have such a $h_1$ resonance. We first
examine the dependence of $\Oh2$ on the specific NMSSM parameters $\l, \k$  in
fig.~\ref{fig:mu700m110_lk}. In this plane, small values of $\k/\l$ are excluded
by either LEP Higgs searches, an unphysical global minimum of the scalar
potential or a negative mass squared for the lightest pseudo-scalar. Large
values of $\k$ give rise to a Landau pole and are also excluded. Most of the
remaining parameter space is allowed by the WMAP constraints.

\begin{figure}[htbp]
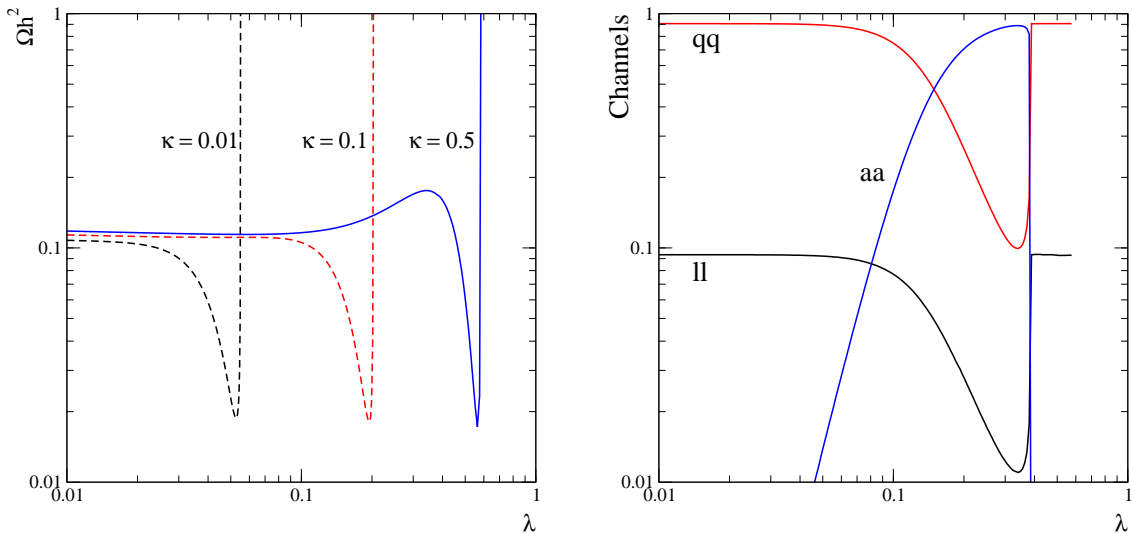

\bce
\begin{picture}(160,75)
\put(0,-7){\mbox{\epsfig{file=fig4a.eps,height=7cm}}}
\put(80,-7){\mbox{\epsfig{file=fig4b.eps,height=7cm}}}
\end{picture}
\ece
\caption{a) $\Oh2$ vs $\l$ for $\mu=700$~GeV,$M_2=111$~GeV,$\k=0.01,0.1,0.5$,
$A_\l=500$~GeV, $A_\k=0$ and $\tb=5$. b) Relative contribution of the main
annihilation channels for the case $\k=0.5$.}
\label{fig:mu700m110_l}
\end{figure}

Next we fix $\k$ and study the variations of $\Oh2$ as a function of $\l$ in
fig.~\ref{fig:mu700m110_l}. We also display the relative contributions of the
most important channels. For $\l=0.01, \k=0.1$, one has $\mnt 1=54$~GeV and
$m_{h_1}=117$~GeV so that $m_{h_1}-2m_{\nt 1}=9$~GeV. A tiny higgsino fraction
($\lsim 10^{-5}$) is then sufficcient to give the right amount of relic density
($\Oh2=0.113$). As $\l$ increases, $m_{h_1}$ decreases whereas $m_{\nt 1}$
remains almost constant. Hence the mass difference decreases and the relic
density drops sharply. When the LSP mass is above the light Higgs resonance
there is no efficient mechanism for LSP annihilation and the value of the relic
density shoots up. Note that for $\k=0.5$, $\Oh2$ first increases with $\l$
before dropping, due to a small increase of the $h_1$ mass with $\l$, and so of
$m_{h_1}-2m_{\nt 1}$. This corresponds to the small region at large $\k$ where
the relic density is above the WMAP limit in fig.~\ref{fig:mu700m110_lk}.

The main annihilation channels are characteristics of the dominant decay modes
of the $h_1$. For small $\l$, the main channels are $\nt 1\nt 1 \to b\bar{b}$
and $\tau\bar{\tau}$. When $\l$ increases, the channel $\nt 1 \nt 1 \to a_1 a_1$
becomes dominant, $a_1$ being a light quasi pure singlet pseudo-scalar. This is
due to the fact that the $h_1a_1a_1$ coupling increases with $\l$, as can be
seen from eq.~(\ref{haacoup}). On the other hand, the $a_1$ mass also increases
with $\l$ (cf. ${\cal M}_{P,33}$ in eq.~(\ref{pseudoscalaire}) with $A_\k=0$),
so that for $\l \gsim 0.4$ this channel becomes kinematically forbidden and the
main channels are again $b\bar{b}$ and $\tau\bar{\tau}$.

\subsection{Annihilation Through Resonances} \label{sec:res}

One of the new features of the NMSSM is its much richer scalar/pseudo-scalar
Higgs sector as compared to the MSSM. Thus, one expects new possibilities for
having $\Oh2 \approx 0.1$ corresponding to annihilation of the LSP through a
Higgs resonance. Starting from a point to the right of the WMAP allowed band in
fig.~\ref{fig:mum2} (where the relic density is large) we will now vary the
Higgs sector parameters $A_\l$, $A_\k$ to see whether it is possible to find
resonances. $A_\l$ directly determines the heavy Higgs doublet mass $m_A$,
eq.~(\ref{mA}), whereas $A_\k$ influences the Higgs singlet masses,
eq.~(\ref{mSP}). The neutralino sector does not depend on these parameters.

\begin{figure}[htbp]
\bce
\begin{picture}(165,54)
\put(0,-7){\mbox{\epsfig{file=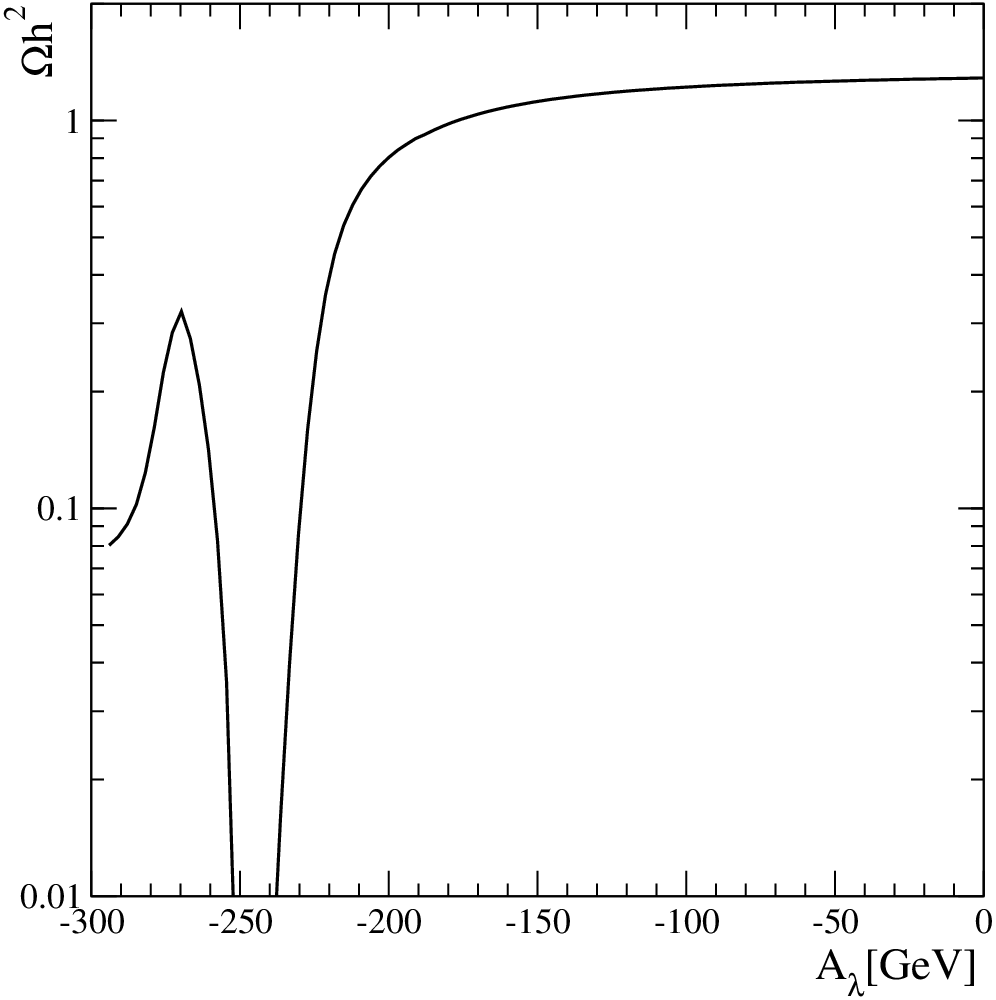,height=4.9cm}}}
\put(55,-7){\mbox{\epsfig{file=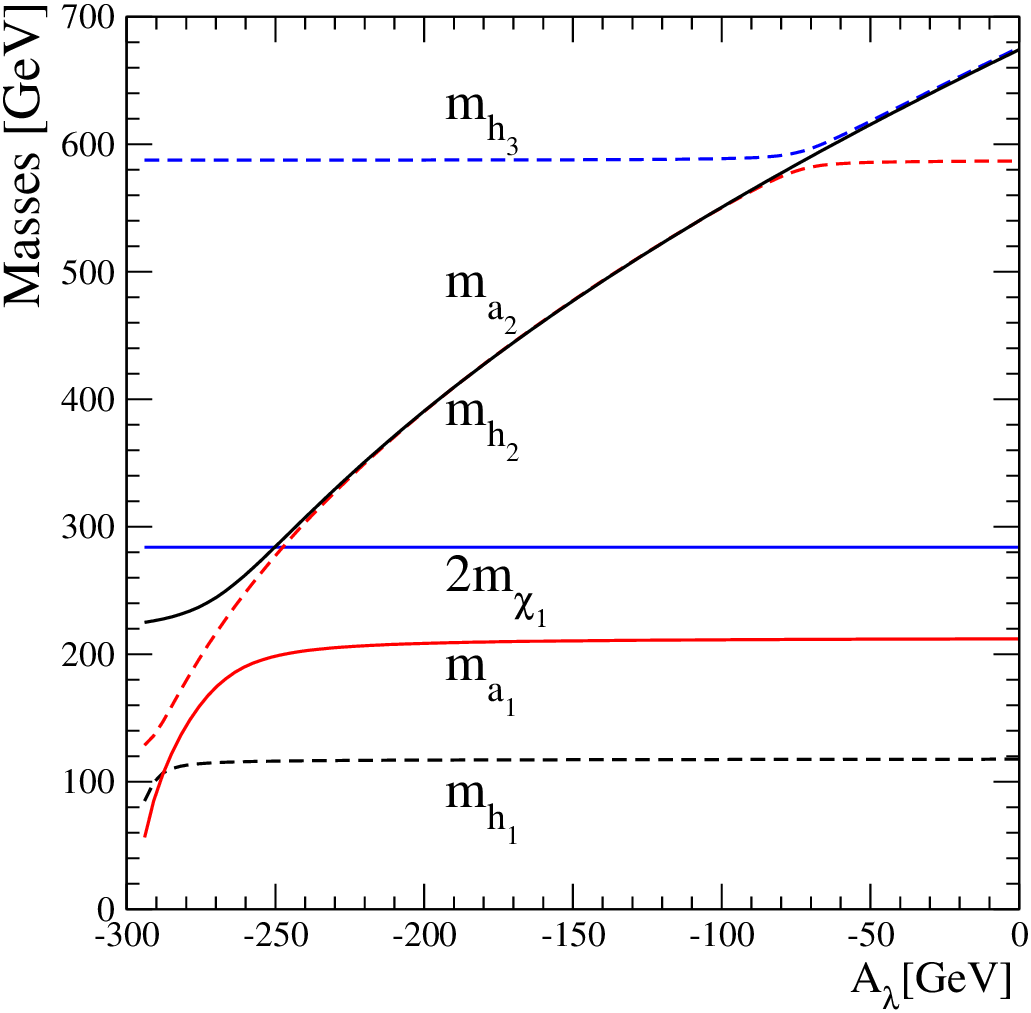,height=4.9cm}}}
\put(110,-7){\mbox{\epsfig{file=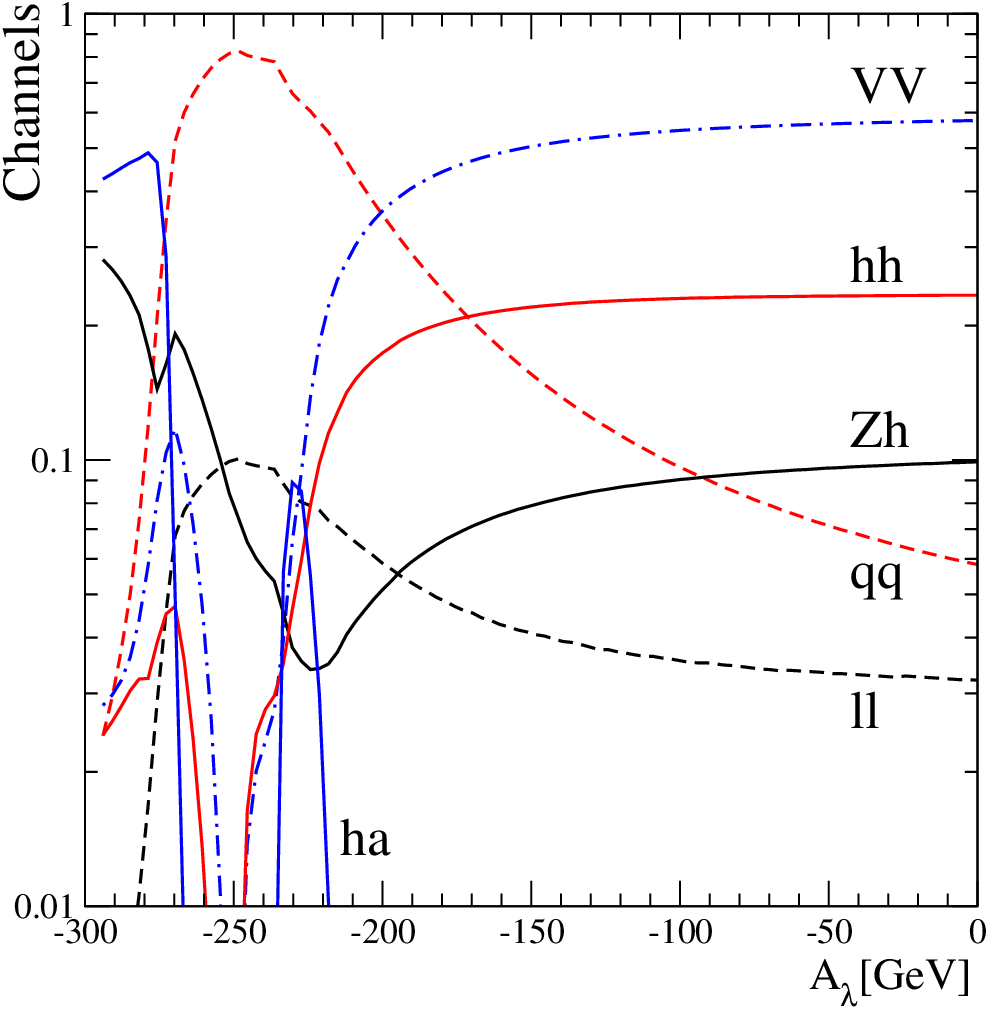,height=4.9cm}}}
\end{picture}
\ece
\caption{$\Oh2$ vs $A_\l$ for $\mu=300$~GeV,$M_2=300$~GeV,$\k=0.1,\l=0.1,
A_\k=-50$~GeV and $\tb=5$. b) Masses of LSP, of scalars (dash) and
pseudo-scalars (full) c) Relative contribution of the main annihilation
channels.} \label{fig:al}
\end{figure}

\begin{figure}[htbp]
\bce
\begin{picture}(165,54)
\put(0,-7){\mbox{\epsfig{file=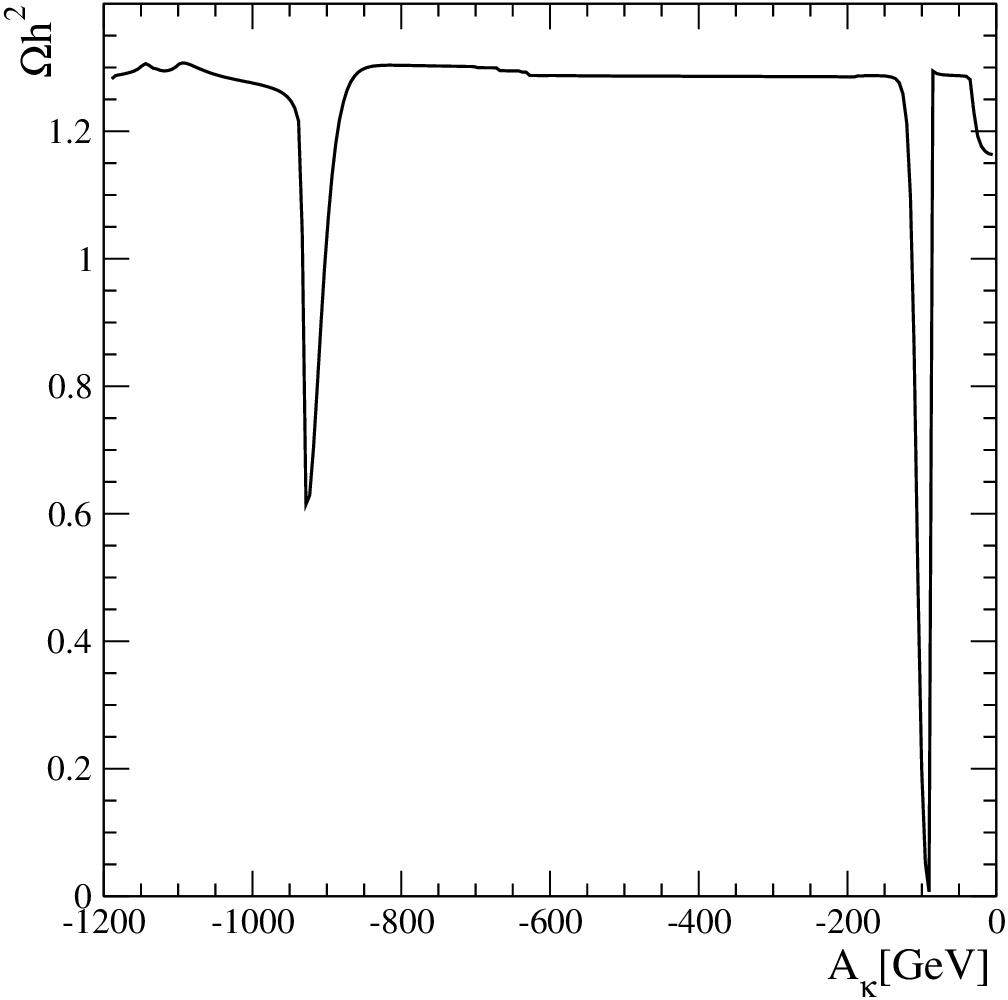,height=4.9cm}}}
\put(55,-7){\mbox{\epsfig{file=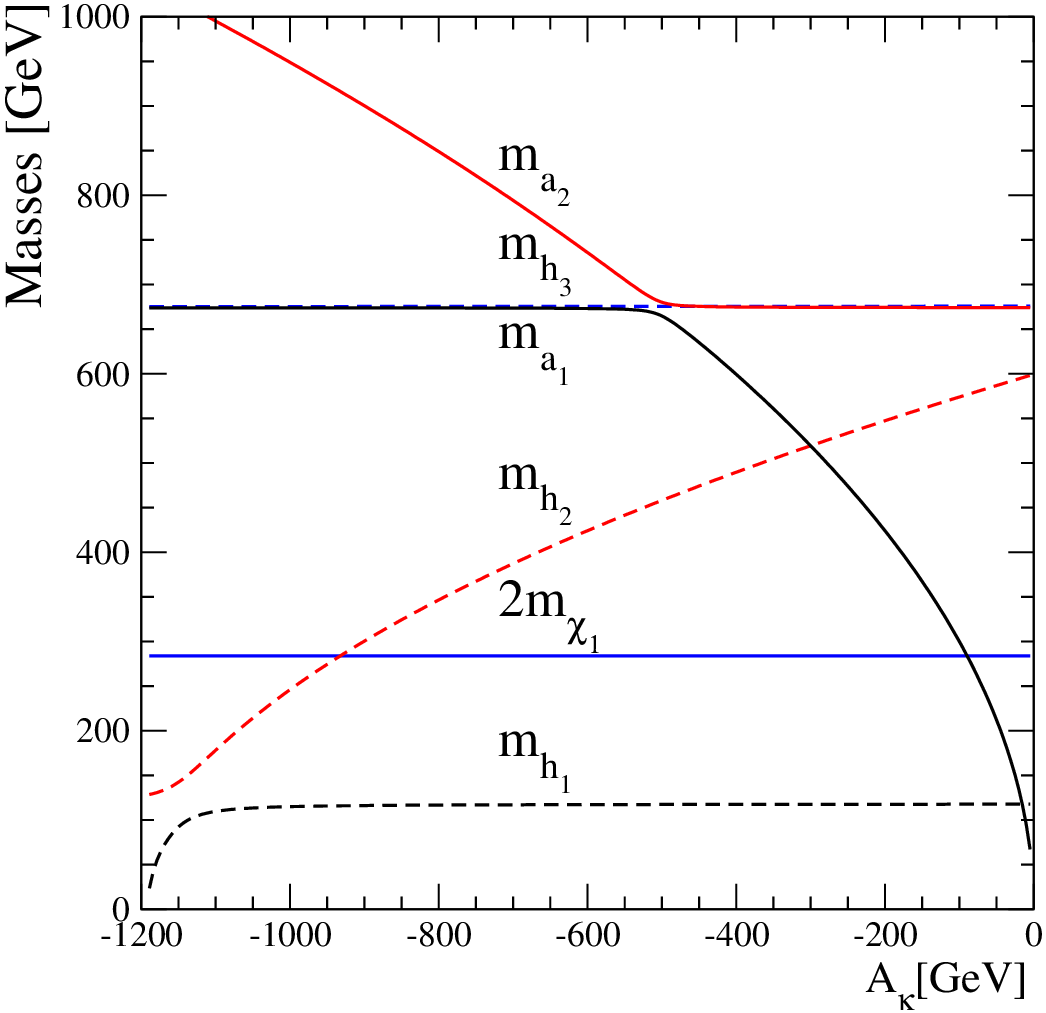,height=4.9cm}}}
\put(110,-7){\mbox{\epsfig{file=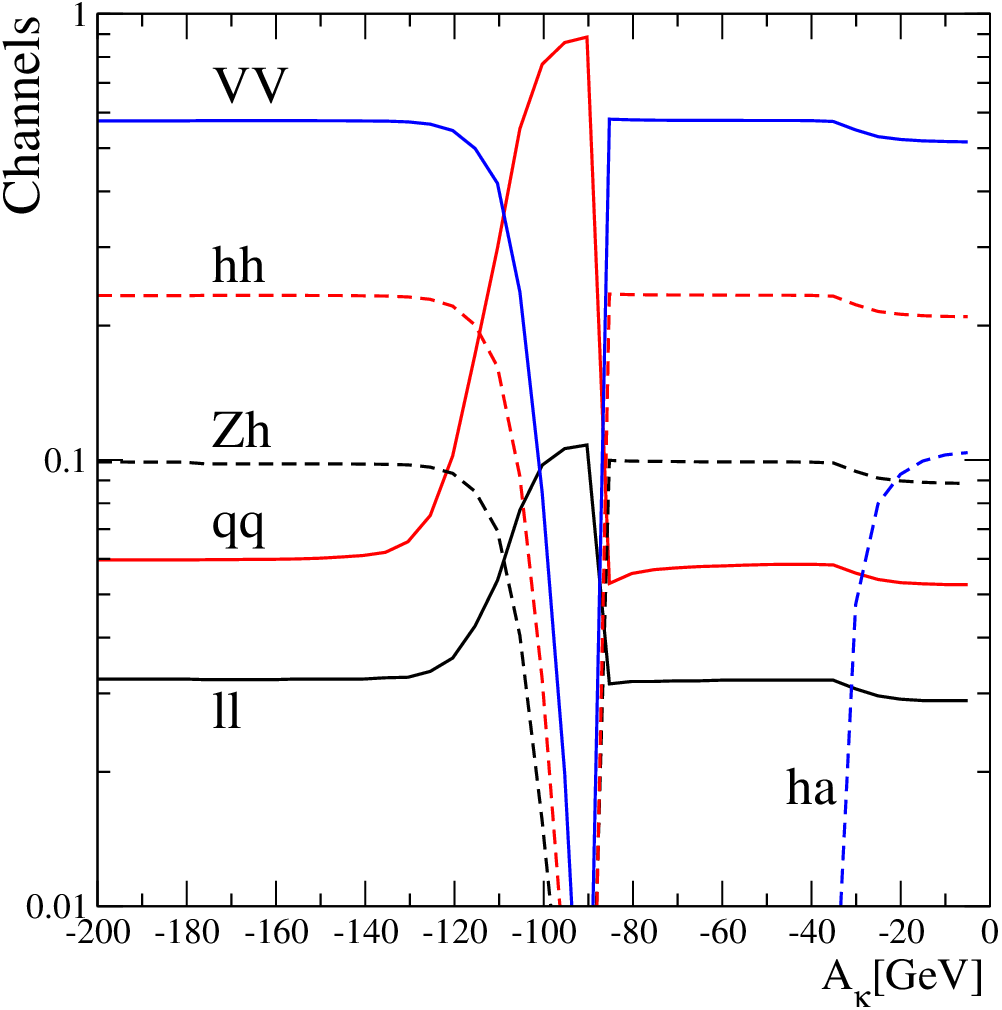,height=4.9cm}}}
\end{picture}
\ece
\caption{$\Oh2$ vs $A_\k$ for $\mu=300$~GeV,$M_2=300$~GeV,$\k=0.1,\l=0.1,
A_\l=0$ and $\tb=5$. b) Masses of LSP, of scalars (dash) and pseudo-scalars
(full) c) Relative contribution of the main annihilation channels.} 
\label{fig:ak}
\end{figure}

First, we fix $\l=0.1$, $\k=0.1$, $\mu=M_2=300$~GeV and $A_\k=-50$~GeV and we
vary $A_\l$. Results for $\Oh2$, the LSP and Higgs masses, as well as the the
main channels contributing to the LSP annihilation are displayed in
fig~\ref{fig:al}. For this choice of parameters the LSP mass is $\mnt
1=142$~GeV  with a bino fraction of 92\%, a higgsino fraction of 7\%, and a
negligible singlino component. Thus the relic density is rather high, $\Oh2=
1.3$ for $A_\l=0$. The main annihilation channel is into gauge boson pairs.
Subdominant channels are into $h_1h_1$, $Zh_1$ as well as fermion pairs. As
$A_\l$ decreases, so does $m_A$ and for $A_\l \approx -250$~GeV, the masses of
the the heavy scalar and pseudo-scalar Higgs doublets, $h_2$ and $a_2$, are such
that $m_{a_2}=m_{h_2}=2\mnt 1$. The relic density then drops sharply. The main
decay modes of both $a_2$ and $h_2$ being $b\bar{b}$ and $\tau\bar{\tau}$, the
dominant LSP annihilation channel near the resonance is into fermion pairs. When
$A_\l\approx -300$~GeV and one moves away from the s-channel resonance, the LSP
still annihilates efficiently into $h_1a_1$ or $Zh_1$ through $a_2$ exchange.
The relic density again falls into the WMAP allowed range. However, in this
region $h_1$ is excluded by LEP.

Next, we fix $A_\l=0$ and vary $A_\k$, keeping all the other parameters as
above. Results for $\Oh2$, the LSP and Higgs masses, as well as the the main
annihilation channels are displayed in fig~\ref{fig:ak}. Note however that the
$A_\l$ range in fig~\ref{fig:ak}c is not the same as in fig~\ref{fig:ak}a-b. The
LSP has the same characteristics as in the previous case. For $A_\k \approx
-950$~GeV, the scalar (singlet) $h_2$ is such that $m_{h_2}=2\mnt 1$, and for
$A_\k \approx -90$~GeV, the pseudo-scalar (singlet) $a_1$ is such that
$m_{a_1}=2\mnt 1$. Far from these resonances, $\Oh2$ is large and the preferred 
annihilation channel is into W pairs. The $h_2$ resonance is associated with an
increase in the annihilation channel $\nt 1\nt 1\to h_1h_1$. However LSP
annihilation through this resonance is not efficient enough to bring the relic
density within the WMAP range. This is because the coupling $\nt 1\nt 1 h_2$,
eq.~(\ref{neuneuh}), is suppressed for a bino LSP and a singlet Higgs. Near the
$a_1$ resonance, the pseudo-scalar annihilates mainly into $b\bar{b}$ as shown
in fig.~\ref{fig:ak}c. Although the $\nt 1\nt 1 h_2$ coupling,
eq.~(\ref{neuneua}), is also suppressed, annihilation is more efficient through
$a_1$ than through $h_2$ because of the p-wave suppression factor for a scalar
exchange.

\subsection{Mixed bino/higgsino: $\mu\approx M_2$ in the NMSSM} \label{sec:mixed}

\begin{figure}[htbp]
\bce
\begin{picture}(90,95)
\put(0,-7){\mbox{\epsfig{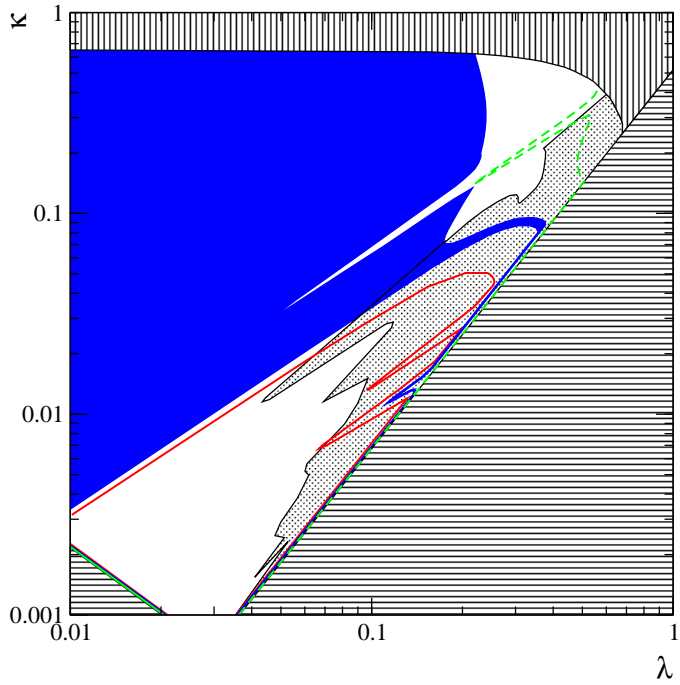}}}
\end{picture}
\ece
\caption{Contour plots for $\Oh2=0.02$ (dash), $\Oh2=1$ (full) and region for
which $0.0945 < \Oh2 < 0.1287$ (blue) in the $\l,\k$ plane for $\mu=220$~GeV,
$M_2=320$~GeV, $A_\l=500$~GeV, $A_\k=0$ and $\tb=5$. Theoretically and
experimentally excluded regions are labeled as in fig.~\ref{fig:mu700m110_lk}.}
\label{fig:mu220m320_lk}
\end{figure}

We now consider in more details the mixed bino/higgsino case with $\mu=220$~GeV,
$M_2=320$~GeV, $\tb=5$, $A_\l=500$~GeV and $A_\k=0$ (corresponding to the second
WMAP allowed region in fig.~\ref{fig:mum2}). In fig.~\ref{fig:mu220m320_lk} we
plot the $\Oh2$ dependence on the specific NMSSM parameters $\l,\k$.
Theoretically and experimentally excluded regions are similar to those obtained
in fig.~\ref{fig:mu700m110_lk}. In the limit $\l \to 0$, the singlino decouples.
According to eq.~(\ref{singmass}), $\mu$ being fixed, the singlino mass is
proportional to $\k/\l$. Hence, for small $\l$ and $\k/\l\ \gsim\ 1/3$, the
singlino is heavy and the LSP is a mixed bino (70\%) and higgsino (30\%), with a
mass $\mnt 1 = 140$~GeV. Its relic density is within the WMAP bounds, the main
annihilation channel being into $WW$, as explained in section~\ref{sec:mum2}.
Smaller values of $\k/\l$ lead to a singlino LSP whose relic density is very
large unless one has recourse to some special annihilation mechanism, as will be
explained in section~\ref{sec:sing}. In the singlino region, one can already see
two clear resonances bringing the relic density down for $\l \approx 0.1$ and
$\k \approx 0.01$. One corresponds to an $h_2$ resonance and the other to a $Z$
resonance. However, these are located in a region ruled out by Higgs searches at
LEP.

\begin{figure}[htbp]
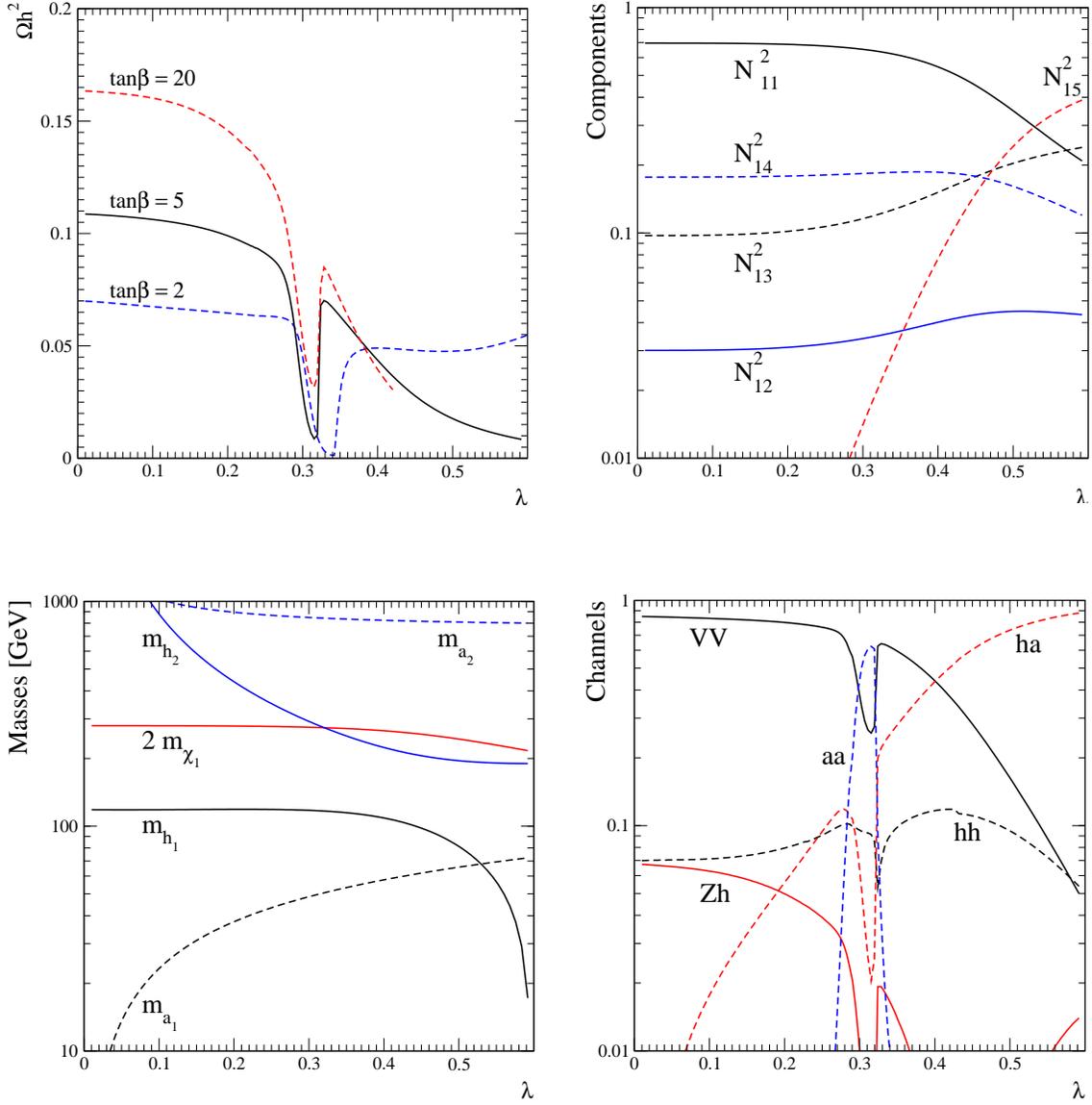

\bce
\begin{picture}(160,155)
\put(0,75){\mbox{\epsfig{file=fig8a.eps,height=7cm}}}
\put(80,75){\mbox{\epsfig{file=fig8b.eps,height=7cm}}}
\put(0,-7){\mbox{\epsfig{file=fig8c.eps,height=7cm}}}
\put(80,-7){\mbox{\epsfig{file=fig8d.eps,height=7cm}}}
\end{picture}
\ece
\caption{a) $\Oh2$ vs $\l$ for $\mu=220$~GeV, $M_2=320$~GeV, $\k=0.2 $,
$A_\l=500$~GeV, $A_\k=0$~GeV, $\tb=5$ (full) and $\tb=2,20$ (dashed). b)
Components of the LSP c) Masses of LSP, of scalars (dash) and of pseudo-scalars
(full) d) Relative contribution of main channels for neutralino annihilation. }
\label{fig:lambda}
\end{figure}

At larger values of $\l$ and $\k$, one can see a third resonance as well as a
decrease of the relic density below the WMAP allowed region. To see more
precisely what occurs in this area of the parameter space, we fix $\k=0.2$ and
plot $\Oh2$ as a function of $\l$ in fig.~\ref{fig:lambda}. We also plot the LSP
components, the LSP and Higgs spectrum, and the main annihilation channels. At
small values of $\l$, one can see from fig.~\ref{fig:lambda}b that the singlino
almost decouples and one recovers the MSSM. The value of $\Oh2$ for $\l \to 0$
and $\tb=5$, fig.~\ref{fig:lambda}a, is equal to the MSSM prediction for
equivalent parameters. The main annihilation channel, fig.~\ref{fig:lambda}d, is
$WW$ through $Z$ exchange, with subdominant $Zh_1$ and $h_1h_1$ contributions
coming from $h_1$ exchange. As $\l$ increases, the relic density decreases until
one encounters the $h_2$ resonance for $\l \approx 0.3$, fig~\ref{fig:lambda}c.
The relic density then drops and the main annihilation channels reflect the
$h_2$ decay modes, with a dominant $a_1a_1$ final state. Above the resonance,
$\Oh2$ keeps decreasing as the LSP becomes dominantly singlino. Recall that for
$\mu$ and $\k$ fixed, the singlino diagonal mass term, eq.~(\ref{neumatrix})
goes like $1/\l$ while the singlino mixings are proportional to $\l$. This
result for the relic density is counter-intuitive, since for a singlino LSP,
annihilation should be suppressed and relic density large. The main annihilation
channel in this case is $h_1a_1$ through t-channel $\nt 1$ exchange. The
decrease of $\Oh2$ with $\l$ can then be understood by an increase of the $\nt 1
\nt 1 h_1$ and $\nt 1 \nt 1 a_1$ couplings as given in
eqs.~(\ref{neuneuh},\ref{neuneua}) for a singlino LSP. However, this area of the
parameter space is excluded by Higgs searches at LEP, as shown in
fig.~\ref{fig:mu220m320_lk}. We will investigate in section~\ref{sec:sing}
whether it is possible to find regions in the parameter space with a singlino
LSP allowed both by LEP and WMAP constraints.

A similar dependence of $\Oh2$ on $\l$ is observed for different values of
$\tb$, in fig.~\ref{fig:lambda}a. The mass of the LSP varies with $\tb$, so the
value of $\l$ for which resonant annihilation occurs is shifted. As we have
discussed in section~\ref{sec:mum2}, when the main annihilation channel is into
gauge boson pairs, annihilation is more efficient at low values of $\tb$ where
the LSP is less bino-like. For $\tb=2$ however, no decrease of $\Oh2$ occurs at
large values of $\l$. This is due to the fact that both the $\nt 1 \nt 1 a_1$
and $\nt 1 \nt 1 h_1$ couplings are much weaker for low values of $\tb$.

\subsection{Singlino LSP} \label{sec:sing}

We explore now scenarios satisfying both LEP and WMAP constraints with a
predominantly singlino LSP. For this we scanned over the whole parameter space
of the NMSSM in the range $\l<0.75$, $\k<0.65$, $2<\tb<10$, $100<\mu<500$~GeV,
$100<M_2<1000$~GeV, $0<A_\l<1000$~GeV and $0<-A_\k<500$~GeV. We found three
classes of models: a mixed singlino/higgsino LSP that annihilates mainly into
$h_1a_1$ and $VV$, an almost pure singlino that annihilates through a $Z$ or Higgs
resonance and a singlino where dominant channels are coannihilation ones. In
table~\ref{tab:singlino} we show a selection of benchmark points along these
lines.

\begin{table}[htbp]
 \begin{center}
 \footnotesize
 \begin{tabular} {|l|r|r|r|r|r|r|}
 \hline
 Point & 1 & 2 & 3 & 4 & 5 & 6\\
 \hline \hline
 $\l$                & 0.6    & 0.24   & 0.4    & 0.23   & 0.31   & 0.0348 \\
 \hline
 $\k$                & 0.12   & 0.096  & 0.028  & 0.0037 & 0.006  & 0.0124 \\
 \hline
 $\tb$               & 2      & 5      & 3      & 3.1    & 2.7    & 5 \\
 \hline
 $\mu$~[GeV]         & 265    & 200    & 180    & 215    & 210    & 285 \\
 \hline
 $A_{\l}$~[GeV]      &  550   & 690    & 580    & 725    & 600    & 50 \\
 \hline
 $A_{\k}$ ~[GeV]     & -40    & -10    & -60    & -24    & -6     & -150 \\
 \hline
 $M_2$~[GeV]         & 1000   & 690    & 660    & 200    & 540    & 470 \\
 \hline \hline
 $\mnt 1$~[GeV]      & 122    & 148    & 35     & 10     & 15     & 203 \\
 \hline
 $N_{13}^2+N_{14}^2$ & 0.12   & 0.29   & 0.12   & 0.03   & 0.06   & 0.02 \\
 \hline
 $N_{15}^2$          & 0.88   & 0.69   & 0.87   & 0.95   & 0.94   & 0.96 \\
 \hline
 $\mnt 2$~[GeV]      & 259    & 199    & 169    & 87     & 182    & 214 \\
 \hline
 $\mch 1$~[GeV]      & 258    & 193    & 171    & 139    & 196    & 266 \\
 \hline
 $m_{h_1}$~[GeV]     & 117    & 116    & 36     & 22     & 34     & 115 \\
 \hline
 $S_{13}^2$          & 0.88   & 0.04   & 0.98   & 1.00   & 1.00   & 0.04 \\
 \hline
 $m_{h_2}$~[GeV]     & 128    & 158    & 117    & 114    & 113    & 163 \\
 \hline
 $S_{23}^2$          & 0.11   & 0.96   & 0.01   & 0.00   & 0.00   & 0.96 \\
 \hline
 $m_{a_1}$~[GeV]     & 114    & 59     & 56     & 18     & 15     & 214 \\
 \hline
 $P_{12}^{'2}$       & 0.99   & 1.00   & 0.99   & 1.00   & 0.99   & 1.00 \\
 \hline \hline
 $\Oh2$              & 0.1092 & 0.1179 & 0.1155 & 0.1068 & 0.1124 & 0.1023 \\
 \hline
 & $ha$ (73\%) & $VV$ (75\%) & $qq$ (65\%) & $qq$ (93\%) & $aa$ (92\%) &
 $\nt 2\nt 2\to X$ (81\%) \\
 & $VV$ (13\%) & $ha$ (17\%) & $ll$ (35\%) & $ll$ (7\%) & $qq$ (7\%) &
 $\nt 1\nt 2 \to X$ (15\%) \\
 Channels
 & $Zh$ (8\%) & $ hh$ (5\%) & & & $ll$ (1\%) & $\nt 1\ch 1 \to X$ (2\%) \\
 & $hh$ (3\%) & $ Zh$ (2\%) & &	& & $qq$ (2\%)\\
 & $qq$ (2\%) & & & & & \\
 & $ll$ (1\%) & & & & & \\
 \hline
 \end{tabular}
 \end{center}
\caption{Benchmark points with a singlino LSP satisfying both LEP and WMAP
constraints}
\label{tab:singlino}
\end{table}

The first scenario is one for which $\mu \ll M_2$ and the LSP is a mixed
higgsino/singlino. We give two different examples in table~\ref{tab:singlino}.
For point 1, the LSP is 88\% singlino and 12\% higgsino, with a mass of 122~GeV.
The main annihilation mode is $h_1a_1$ through t-channel $\nt 1$ exchange, $h_1$
and $a_1$ being both mainly singlet (88\% and 99\% respectively). As we have
seen in section~\ref{sec:mixed} this is due to the enhanced couplings $\nt 1 \nt
1 a_1(h_1)$ for large values of $\l$. The annihilation into $h_1a_1$ being too
efficient for large values of $\tb$, we chose $\tb=2$ for this point.
Annihilation of the higgsino component into W pairs accounts for the subdominant
channel. For point 2, the LSP mass is 148~GeV and the main annihilation channel
is $WW$, which necessitates some sizeable higgsino component (29\% here). The
subdominant channel $h_1a_1$ implies the singlino component of the LSP (69\%).
The singlino fraction cannot be maximal here since there are no important
annihilation channels for a pure singlino when $\l$ is not so large and the
channel $h_1a_1$ is not dominant.

We give in table~\ref{tab:singlino} the characteristics of three typical
scenarios with light singlinos. The only efficient annihilation mode for a very
light singlino (below 50~GeV) is via a $Z$ or a Higgs resonance. For point 3,
the LSP is 87\% singlino with a mass of 35~GeV and annihilates through a $Z$
exchange. In this scenario, although the singlino component dominates, a 12\%
higgsino component is sufficient to ensure efficient annihilation through the
$Z$. As for the main annihilation channels they are characteristic of the $Z$
decay modes, mainly into quark pairs and lepton pairs. Neutralinos of $30$~GeV
that satisfy the WMAP upper bound were also found in SUGRA models with
non-universal gaugino masses~\cite{LSP_light}. Because the scalar/pseudo-scalar
Higgs states in the NMSSM can be much below the $Z$ mass while passing all the
LEP constraints, we expect to find even lighter singlino LSP's in models where
$2 m_{\nt 1}\approx m_h$. We show two examples (point 4 and 5) of such models in
table~\ref{tab:singlino}. In both cases the LSP annihilates via a light scalar
dominantly singlet ($S_{13}^2 \approx 1$). This scalar decays either into
$b\bar{b}$ (point 4) or, when kinematically accessible, into $a_1a_1$ (point 5),
the $a_1$ being also mainly singlet. The Higgs sector of such models is of
course severely constrained by LEP, in particular the limit on the SM-like
scalar, here the second scalar, $h_2$. For this reason most scenarios with light
singlinos have $\tb\approx 3$ which is the value for which the lightest visible
({\it ie} non singlet) Higgs mass, $m_{h_2}$, is maximized~\cite{NHIGGS}. Note
that a light singlino requires $\k \ll \l$ and not too large value for $\mu$.
The singlino masses for point 4 and 5 are respectively 10~GeV and 15~GeV. We
found points in the parameter space with singlinos as light as a few GeV.

For $\k\ \lsim\ \l \ll 1$, the LSP is heavy with a large singlino component. No
efficient annihilation mechanism is then available. However coannihilation with
heavier neutralinos and charginos can be very efficient especially for a
higgsino-like NLSP. Point 6 in table~\ref{tab:singlino} gives an example of such
a scenario. The LSP is 96\% singlino with a mass of 204~GeV. The mass difference
with the NLSP $\nt 2$ is $11$~GeV. The coannihilation channels are
overwhelmingly dominant. The $\nt 2$ higgsino component is just enough (28\%)
for efficient annihilation. The main channels are $\nt 2 \nt 2 (\nt 1) \to
t\bar{t}, b\bar{b}$ and correspond to annihilation through $h_3$ and $a_2$
exchange. For this point, $h_3$ and $a_2$ belong to the heavy Higgs doublet with
$m_A \approx 470$~GeV, so that we are close to a (double) resonance. Such a
resonance is not necessary though, in order to have efficient $\nt 2$
annihilation. We also found points in the parameter space with a heavy singlino
where the dominant channel was $\nt 2 \nt 2 (\nt 1) \to VV$ through $Z$
exchange.

\section{Conclusions}

In the NMSSM, the same mechanisms as for the MSSM are at work for neutralino
annihilation: into fermion pairs through s-channel exchange of a $Z$ or Higgs,
into gauge boson pairs through either $Z$/h s-channel exchange or t-channel
exchange of heavier neutralinos or charginos. The new feature of the NMSSM is
the presence of additional Higgs states, which means additional regions of
parameter space where rapid annihilation through a s-channel resonance can take
place. We found that annihilation through a Higgs resonance is dominant in large
regions of the parameter space and this even at low to intermediate values of
$\tb$. Furthermore, new channels also open up since light Higgs state can be
present. For example annihilation channels into $Zh,hh,ha$ or even $aa$ can
contribute significantly to the relic density. These proceed through s-channel
$Z$/h/a exchange or t-channel neutralino exchange. Despite the additional
annihilation channel available because of a richer Higgs sector, annihilation of
neutralinos is not always favoured in the NMSSM. In general the singlino
component of the LSP tends to reduce the annihilation cross-section. Therefore
one expects that the relic density of dark matter strongly constrain models with
a large singlino component. We found however regions of the parameter space
where a singlino LSP gives the right amount of dark matter, either for large
$\l$, s-channel resonances into a $Z$ or a Higgs, or coannihilation with $\nt
2$, $\ch 1$.

For a bino LSP, compatibility with the WMAP result can be recovered in the
NMSSM. This however requires tuning the parameters of the model such that
$2m_{\nt 1}$ is only a few GeV below the mass of a Higgs boson. In models where
$M_2=2M_1$ as we have discussed here, this would also
mean that one has a chargino mostly wino with a mass $\mch 1\approx 2m_{\nt
1}\approx m_{h_1}$. Such a degeneracy between the chargino mass and the Higgs
mass could be measured  at the ILC.
If the Higgs boson is mostly doublet, it might be hard to disentangle the NMSSM
from the MSSM. If it is singlet, mass relation between the NMSSM Higgs states
would be different from what expected in the MSSM. A mostly singlet Higgs state
could eventually be observable at the LHC with high luminosity~\cite{Moretti}.

In the case of a mixed bino/higgsino LSP, annihilation relies, as in the MSSM,
on s-channel $Z$ exchange and t-channel neutralino/chargino exchange. If the
singlino state is not heavy and decouples, {\it ie} $\l$ not too small and
$\k$ not too large, the five neutralino states might be visible at the LHC/ILC.
This would be a clear signature of the NMSSM.

Finally, in the singlino LSP case, $\mu$ cannot be too large. One therefore
would expect visible higgsinos at the LHC. The singlino LSP would however appear
at the end of the decay chain in any sparticle pair production process, which
might complicate the detection task as it was the case at LEP~\cite{NLEP}.

As a concluding remark, one should mention that, if one assumes minimal flavour
structure, $b \to s\g$ might impose strong constraints on the NMSSM parameter
space, especially for large values of $\l$ where the charged Higgs mass,
eq.~(\ref{charge}), could be lighter than in the MSSM. A complete study of the
constraints coming from flavour physics should be the next step of the
phenomenological study of the NMSSM.

\section*{Acknowledgements}

This work was supported in part by GDRI-ACPP of CNRS and by grants from the
Russian Federal Agency for Science, NS-1685.2003.2 and RFBR 04-02-17448. The manuscript was finished
in Les Houches, Physics at TeV Colliders 2005.

\newpage

\end{document}